\documentclass[%
 aip,jcp,
 amsmath,amssymb,
 reprint,%
]{revtex4-1}

\usepackage{graphicx}
\usepackage{dcolumn}
\usepackage{bm}

\usepackage[utf8]{inputenc}
\usepackage[T1]{fontenc}
\usepackage{mathptmx}

\usepackage{subfigure}

\begin{document}

\preprint{}

\title[MD simulations of Arg$_9$]{Molecular dynamics studies of interactions between
Arg$_9$(nona-arginine) and a DOPC/DOPG(4:1) membrane}

\author{Seungho Choe}
 \email{schoe@dgist.ac.kr}
\affiliation{$^1$
School of Undergraduate Studies, College of Transdisciplinary Studies, Daegu Gyeongbuk Institute of
Science \& Technology, 333 Techno Jungang-daero, Hyeonpung-eup, Dalseong-gun, Daegu 42988, South Korea
}%


\begin{abstract}
It has been known that the uptake mechanisms of cell-penetrating peptides(CPPs) depend on the
experimental conditions such as concentration of peptides, lipid composition, temperature, etc.
In this study we investigate the temperature dependence of the penetration of Arg$_9$s into a
DOPC/DOPG(4:1) membrane using molecular dynamics(MD) simulations at two different temperatures, T = 310
K and T = 288 K.
Although it is difficult to identify the temperature dependence because of having only one single
simulation at each temperature and no evidence of translocation of Arg$_9$s across the membrane at both
temperatures, our simulations suggest that followings are strongly correlated with the penetration of
Arg$_9$s: a number of water molecules coordinated by Arg$_9$s, electrostatic energy between Arg$_9$s and
the lipids molecules.
We also present how Arg$_9$s change a bending rigidity of the membrane and how a collective behavior
between Arg$_9$s enhances the penetration and the membrane bending. Our analyses can be applicable to
any cell-penetrating peptides(CPPs) to investigate their interactions with various membranes using MD
simulations.
\end{abstract}

\maketitle


\section{\label{sec_intro}Introduction}

Cell penetrating peptides(CPPs), typically comprising 5–30 amino acids, can translocate across cell
membranes. They have been extensively studied for a couple of decades since they are capable of
transporting various cargoes(e.g., proteins, peptides, DNAs, and drugs) into cells,\cite{Guidotti2017}
However, the uptake mechanisms of CPPs are still controversial. It is known that the translocation
mechanisms of different families of CPPs are not the same, and most CPPs can have more than a single
pathway depending on the experimental conditions.\cite{Madani2011} There are two major uptake mechanisms,
i.e., the endocytotic(energy-dependent) pathway and the direct(energy-independent) translocation. The
detailed uptake mechanisms can be found in Refs. \onlinecite{Madani2011,Guidotti2017,Ruseska2020}.

The uptake mechanisms of CPPs depend on various factors such as the physical and chemical properties of CPPs, concentration of CPPs, the properties of the membrane, etc. \cite{Ruseska2020} The concentration of CPPs is one of key factors which gives different uptake pathways. It is generally accepted in many CPPs that
endocytosis occurs at low concentration while the direct translocation occurs at low concentration. However, penetratin shows the reverse. Endocytosis occurs at high concentration while the direct translocation occurs at low concentration in the case of penetratin. \cite{Pae2014} Therefore, the uptake mechanism seems much more complicated than we thought. There are various models to explain the
direct translocation, such as the carpet-like model \cite{Pouny92}, pore formation \cite{Matsuzaki96},
inverted micelle formation \cite{Derossi96}, the membrane thinning model \cite{Lee05}, etc. Detailed
mechanisms of those models are also given in Refs. \onlinecite{Madani2011,Ruseska2020}.

Among various CPPs, arginine-rich peptides seem very interesting.
As mentioned above the switch between different uptake mechanisms might depend on the concentration of the
peptides. It has been shown in experiments that endocytosis is only dominant to a specific low peptide
concentration. \cite{Fretz07}
One of possible scenarios for the
 insertion of arginine-rich peptides into the lipid bilayer is a strong interaction with negatively charged phospholipid heads (see, e.g., Ref. \onlinecite{Ruseska2020}). The charge neutralization results in the membrane deformations and this can
 lead to a decrease in a bending rigidity of the membrane.

 The rapid cytoplasmic entry has been explained
 by the accumulation of the peptides at certain membrane areas, called nucleation
 zones.\cite{Wallbrecher17}
It was also reported that a rapid temperature decrease from 37$^\circ$C to 15$^\circ$C induces an
efficient entry of Arg$_9$(called R9 or nona-arginine in different literature) even at a low peptide
concentration.\cite{Melikov2015} This is why two temperatures(T = 310 K and 288 K) were chosen in our
simulations. Although the mechanism of temperature dependence shown in Ref. \onlinecite{Melikov2015} is
closely related to the calcium signalling, we focus primarily on the temperature dependence of
interactions between Arg$_9$ and the lipids molecules.

It was shown by Eir\'{i}ksd\'{o}ttir et al. \cite{Eiriksdottir2010} that several CPPs can adopt a specific
secondary structure in stabilizing their interactions with membranes. Among them Arg$_9$ exhibited a
random coil in both water and membrane environments, and it means that Arg$_9$ was unable to adopt any
specific secondary structure in both environments.
Eir\'{i}ksd\'{o}ttir et al. \cite{Eiriksdottir2010} also carried out leakage tests to assess the
consequence of interactions between Arg$_9$ and membranes on membrane stability, and found that Arg$_9$
did not induce any leakage, suggesting no membrane perturbation and probably no insertion into the
bilayer.

It was suggested that the entry of Arg$_9$ into DOPG/DOPC-GUV(GUV: the giant unilamella vesicle) and
DLPG/DTPC(2/8)-GUV occurs through transient hydrophilic prepores. \cite{Islam2018} A prepore is an area of
decreased density of lipids due to the fluctuations of local lateral density. Prepores immediately close
because of the large line tension. The interaction of CPPs with these
prepores may stabilize the prepores by decreasing their line
tension. More detailed physical views on prepores and comparison of the line tension of several lipid
mixtures are given in Ref. \onlinecite{Islam2018} and references therein.

Herce et al. \cite{Herce2009} showed in their molecular dynamics(MD) simulations that Arg$_9$ follows
the same mechanism previously found for Tat \cite{Herce2007} by the same authors.
The attractive interactions between the Arg$_9$ peptides and the phosphate groups of the phospholipids results in large local distortions of the bilayer, and these distortions lead to the formation of a toroidal pore. Note that the authors used the system containing a single phospholipid(DOPC) in their MD simulations. However, the authors showed in experiments that anionic phospholipids such as DOPG lipids are not required
for the Arg$_9$ peptides to penetrate bilayers.\cite{Herce2009}

MD simulations have been used to investigate functional properties of Arg$_9$ and its interactions with
many different lipids, however, only a few simulations used mixtures of lipids(e.g., a mixture of
DOPC/DOPG or DOPC/DOPE) so far and the mechanism of translocation of Arg$_9$ and interactions with lipids
are still not conclusive.
It was found that
a DOPC/DOPG membrane appeared the better host for the translocation of KR9C in MD
simulations.\cite{Crosio2019}

In our study we investigate the penetration of Arg$_9$ peptides into a model membrane(e.g., a mixture of
DOPC/DOPG(4:1)) using MD simulations at two different temperatures(T = 310 K and T = 288 K). Our first
goal is to see if Arg$_9$ can translocate across the DOPC/DOPG mixture by making a pore or a prepore, and
the second goal is to investigate the temperature dependence of interactions between Arg$_9$ and the
lipids mixture.

\section{\label{sec_methods}Simulation Methods}

All simulations were performed using the NAMD package \cite{namd} and CHARMM36 force field.\cite{charmm}
CHARMM-GUI \cite{charmm-gui} was used to setup a DOPC/DOPG(4:1) mixture and TIP3P water molecules. The
lipids mixture consists of 76 DOPC and 19 DOPG lipids in each layer. K and Cl ions were added to make a
concentration of 150 mM. One Arg$_9$ peptide was made by the molefacture plugin in VMD \cite{vmd} and it
was duplicated to make four Arg$_9$ peptides. The total number of Arg$_9$ peptides was the same as in
Herce et al.'s \cite{Herce2009} to make a similar concentration of Arg$_9$s. All Arg$_9$ peptides were
initially located in the upper solution and they were bound to the upper layer during the equilibration.
The total number of water molecules after making the system neutralized was 9643. The NPT simulations were
performed at two different temperature T = 288K and 310K in order to investigate the temperature
dependence of penetration of Arg$_9$s. Temperature and pressure were kept constant using Langevin
dynamics. An external electric field(0.05 V/nm) was applied in the negative z direction(from the Arg$_9$s
peptides to the membrane) as suggested in previous work \cite{Herce2009,Walrant2012} to account for the
transmembrane potential. \cite{Roux2008} The particle-mesh Ewald(PME) algorithm was used to compute the
electric forces and the SHAKE algorithm was used to allow a 2 fs time step during the whole simulations.
At the first stage, the system was equilibrated at T = 310 K for 500 ns long. At the second stage, two
systems were prepared: One was a system with the same temperature(T = 310 K) as in the equilibration at
the first stage. The other was a system with T = 288 K. When the temperature was down from T = 310 K to T
= 288 K in the second system, the system was quickly equilibrated at T = 288 K. Both systems(T = 310 K and
288 K) were ran for another 500 ns. The first 200 ns data in both simulations were discarded for the
equilibration and only the last 300 ns data were used for the analyses.

\section{\label{sec_results}Simulation Results}

\subsection*{A number of water molecules coordinated by Arg$_9$s is strongly correlated with the
penetration depth}

We don't see any translocation of Arg$_9$ across the DOPC/DOPG bilayer during the simulations. However, it
can penetrate into the lipids mixture along with a quite number of water molecules. We define a
penetration depth as a distance in the z direction between the center of mass of each Arg$_9$(taking only
C$_\alpha$s for the sake of simplicity) and that of the upper leaflet. Here, a leaflet means a surface
which consists of 95 phosphorus(P) atoms in the upper or the lower layer.

\begin{figure}[ht]
\centering

\includegraphics[width=8cm, height=6cm]{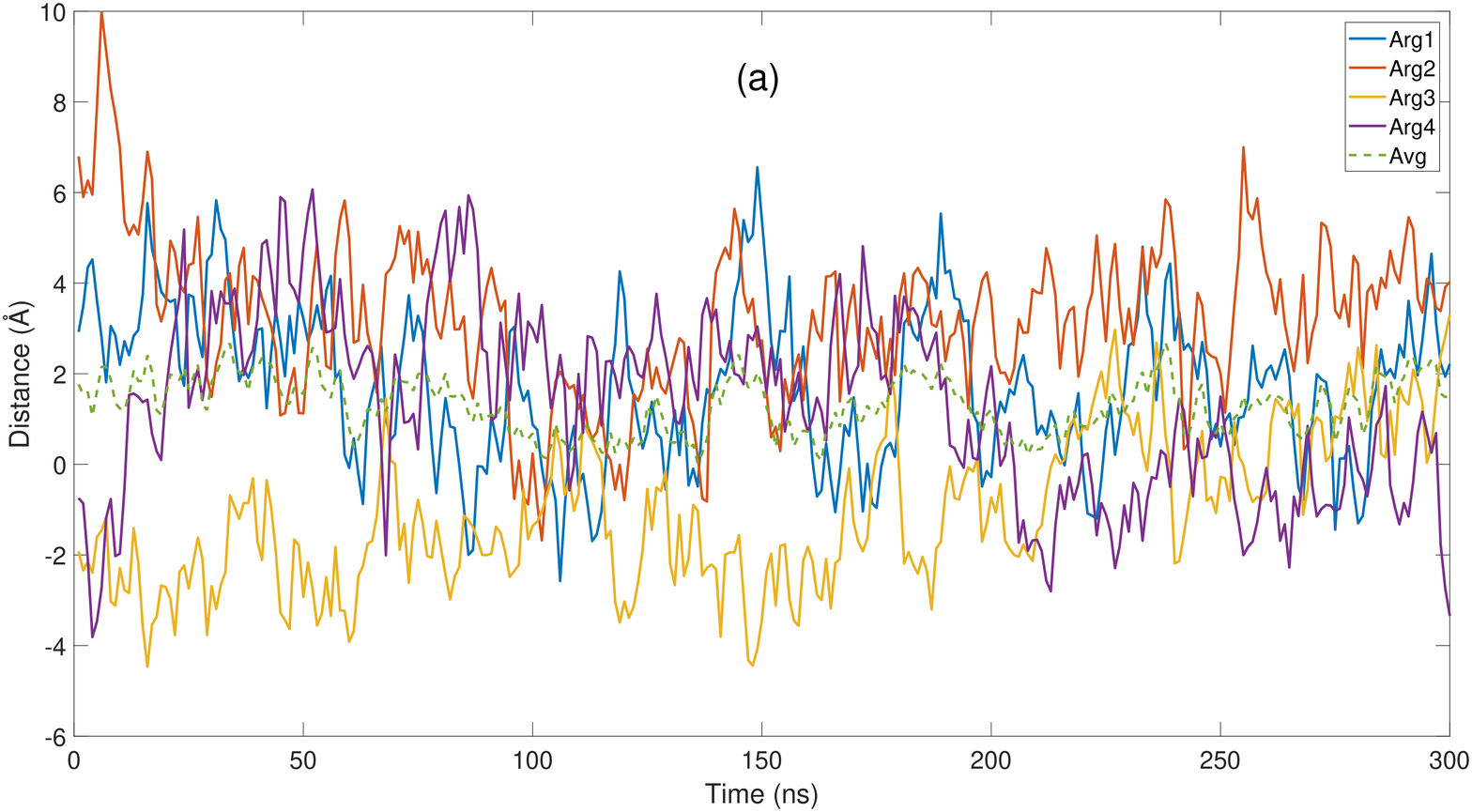}
\includegraphics[width=8cm, height=6cm]{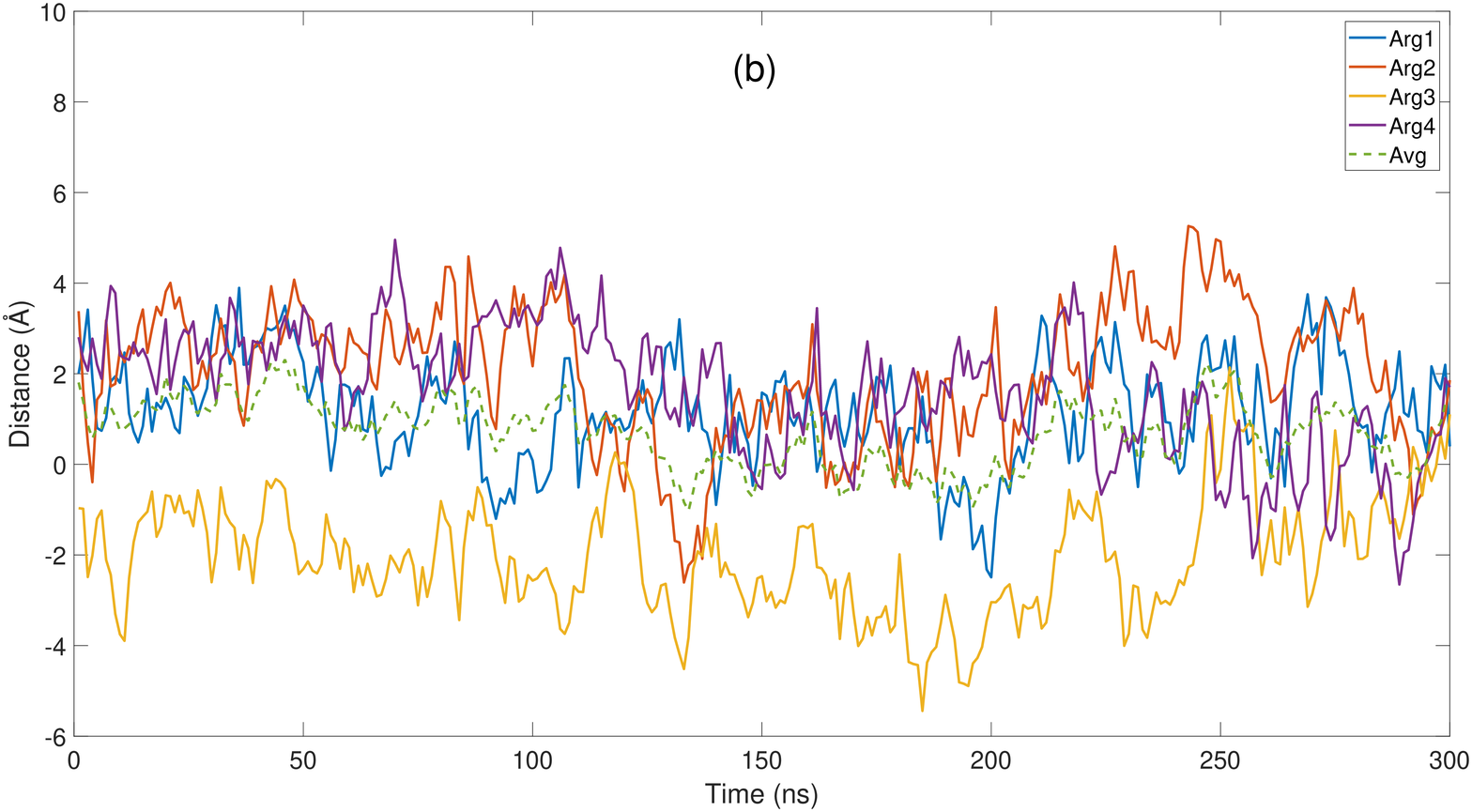}

\caption{The distance between the center of mass of each Arg$_9$ and the center of masses of phosphorus(P)
atoms in the upper leaflet(in the z direction) at (a) T = 310 K (b) T = 288 K, respectively.  The color
code represents each Arg$_9$. Avg(green dotted line) means an average value over four Arg$_9$s.}
\label{fig:com}
\end{figure}

Fig. \ref{fig:com}(a) shows the penetration depth of each Arg$_9$ at T = 310 K.
The solid lines denote the penetration depth of each Arg$_9$, and the dotted line represents an average
over those four values. The positive value means that the center of mass of Arg$_9$ is located above the
center of masses of phosphorus(P) atoms, and the negative means below the phosphorus(P) atoms. One of
Arg$_9$s(Arg3) is located below the upper leaflet during the most of simulation time.

\begin{figure}[ht]
\centering

\includegraphics[width=6cm, height=6cm]{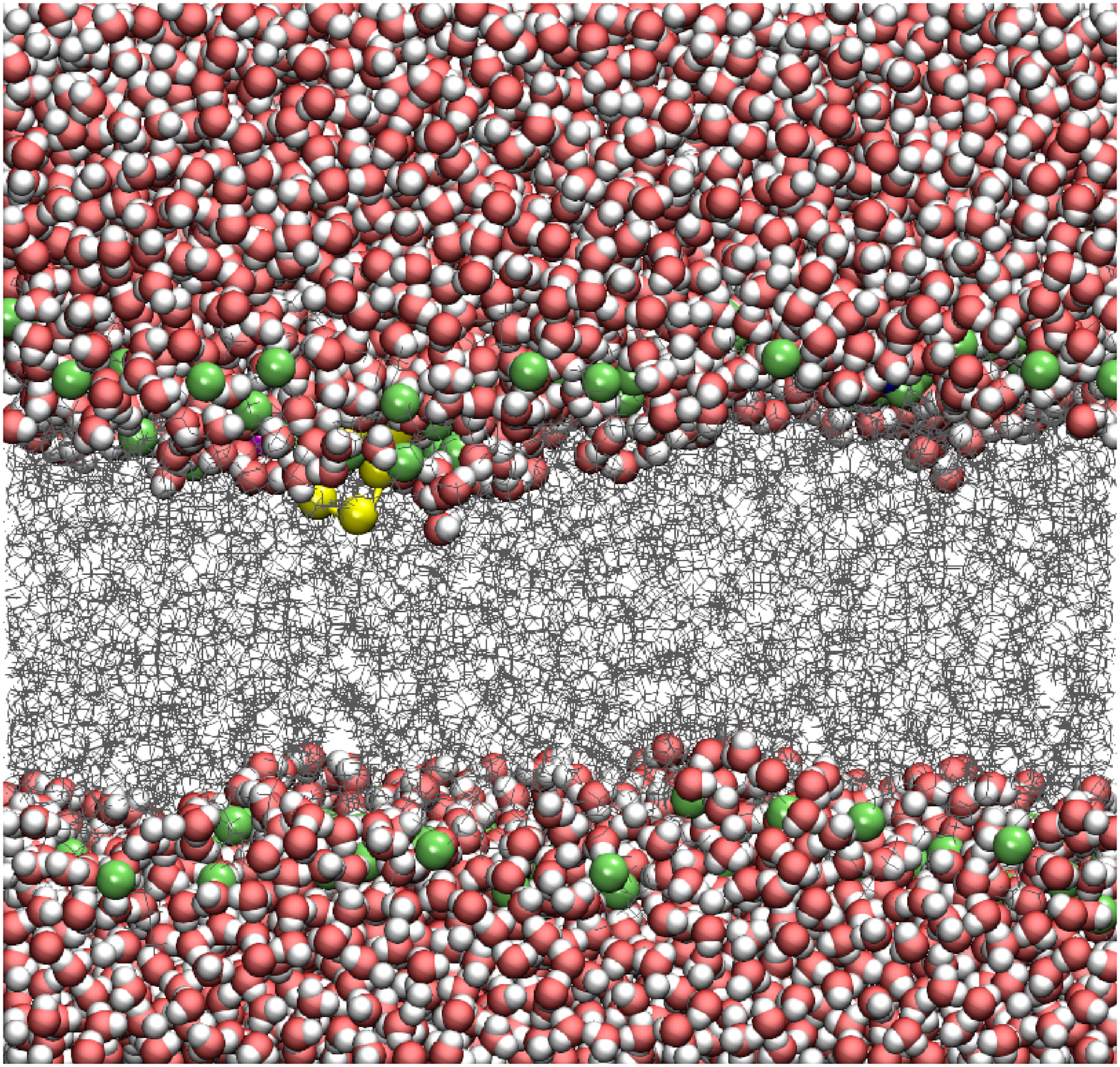}

\caption{A snapshot, which shows the largest penetration depth of Arg3(yellow), at T = 288 K. Only
Arg3(yellow), the phosphorus(P) atoms(green dots), the lipids molecules(grey lines), and water molecules
in vdW of VMD are shown}
\label{fig:K288_38468_snapshots}
\end{figure}

\begin{figure}[ht]
\centering

\includegraphics[width=8cm, height=6cm]{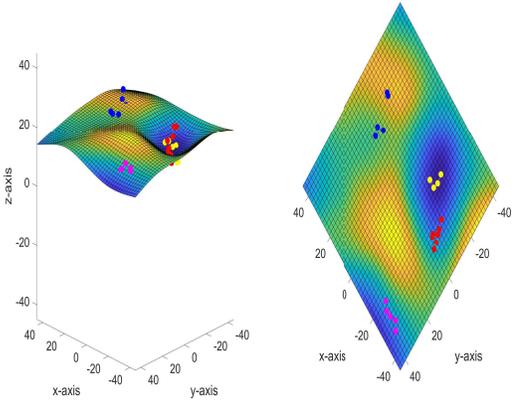}

\caption{Membrane curvature and a relative position of each Arg$_9$(showing C$\alpha$s only) from the
snapshot in Fig. \ref{fig:K288_38468_snapshots}. Representation in 3D(left) and in the x-y plane(right),
respectively. }
\label{fig:K288_38468_curvature}
\end{figure}

Fig. \ref{fig:com}(b) shows the same quantity as in Fig. \ref{fig:com}(a), but at T = 288 K. It shows
that Arg$_9$ can penetrate slightly more into the lipids mixture at T = 288 K.
One of Arg$_9$s(Arg3) is below the upper leaflet during the whole simulation time, and it reaches even -5
{\AA} in this figure. Note that each point in the figures corresponds to a block average over 1 ns. We
find that the largest penetration depth is -6.3 {\AA} during the simulation at T = 288 K, and Fig.
\ref{fig:K288_38468_snapshots} shows a snapshot at that moment. C$_\alpha$s of Arg3 are shown as the same
color code in Fig. \ref{fig:com}. The green dots by the vdW representation of VMD correspond to
phosphorus(P) atoms in the DOPC/DOPG mixture. The lipids molecules are shown as gray lines and water
molecules are represented by vdW of VMD. Fig. \ref{fig:K288_38468_curvature} represents a deformation of
the upper leaflet due to the presence of Arg$_9$s and a relative distance between the upper leaflet and
each Arg$_9$ using the snapshot in Fig. \ref{fig:K288_38468_snapshots}. The lower leaflet is not shown in
this figure for clarity. The blue color in the leaflet denotes a trough and the yellow a crest. Some of
C$\alpha$s located below the leaflet are not seen in this figure.

\begin{figure}[ht]
\centering

\includegraphics[width=8cm, height=6cm]{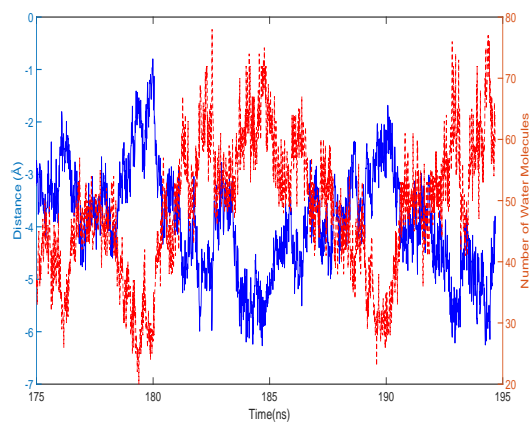}

\caption{The distance between the center of mass of Arg3 and the center of masses of 95 phosphorus(P)
atoms in the upper leaflet(solid blue line, the left axis) and a number of water molecules within 15 {\AA}
of Arg3(dotted red line, the right axis) at T = 288 K. We count only water molecules below the center of
masses of P atoms.}
\label{fig:K288_num_water}
\end{figure}

A quite number of water molecules also penetrate into the DOPC/DOPG mixture and this results in a membrane
thinning and bending. Fig. \ref{fig:K288_num_water} shows  the total number of water molecules located 15
{\AA} apart from the center of mass of Arg3 at T = 288 K. We count only water molecules below the center
of masses of phosphorus(P) atoms. Fig. \ref{fig:K288_num_water} shows a strong correlation between the
penetration depth of Arg$_9$ and the number of water molecules surrounding it. The number of water
molecules increases as the penetration of Arg3 increases. This confirms the previous results from MD
simulations. \cite{Freites2005,Dorairaj2007,MacCallum2007}

At the current stage we can not conclude that the lowering temperature will really help Arg$_9$s to make
more penetration into the lipids mixture. First of all, the difference in average penetration depths at
two temperatures is not that significant(about 1 or 2 {\AA} in average). Second, we have only one single
simulation at each temperature, and the total simulation time(300 ns) may not be long enough to see the
temperature dependence. However, it turns out that the penetration depths are affected by electrostatic
energy between Arg$_9$s and the lipids mixture and we want to focus on how Arg$_9$s interact with the
lipids mixture in the following analyses.

\subsection*{Electrostatic energy between Arg$_9$s and the DOPC/DOPG mixture is important for the
penetration}

It is difficult to imagine how Arg$_9$s can pass through the DOPC/DOPG mixture because DOPG is a
negatively charged lipid and therefore one can expect strong attractive interactions between Arg$_9$s and
DOPG lipids. It was suggested that the electrostatic interaction between positive charges of Arg$_9$s and
negative charges of lipid molecules is crucial at the early stage of translocation.\cite{Walrant2012} We
want to investigate this feature in our simulations.

\begin{figure}[ht]
\centering

\includegraphics[width=8cm, height=6cm]{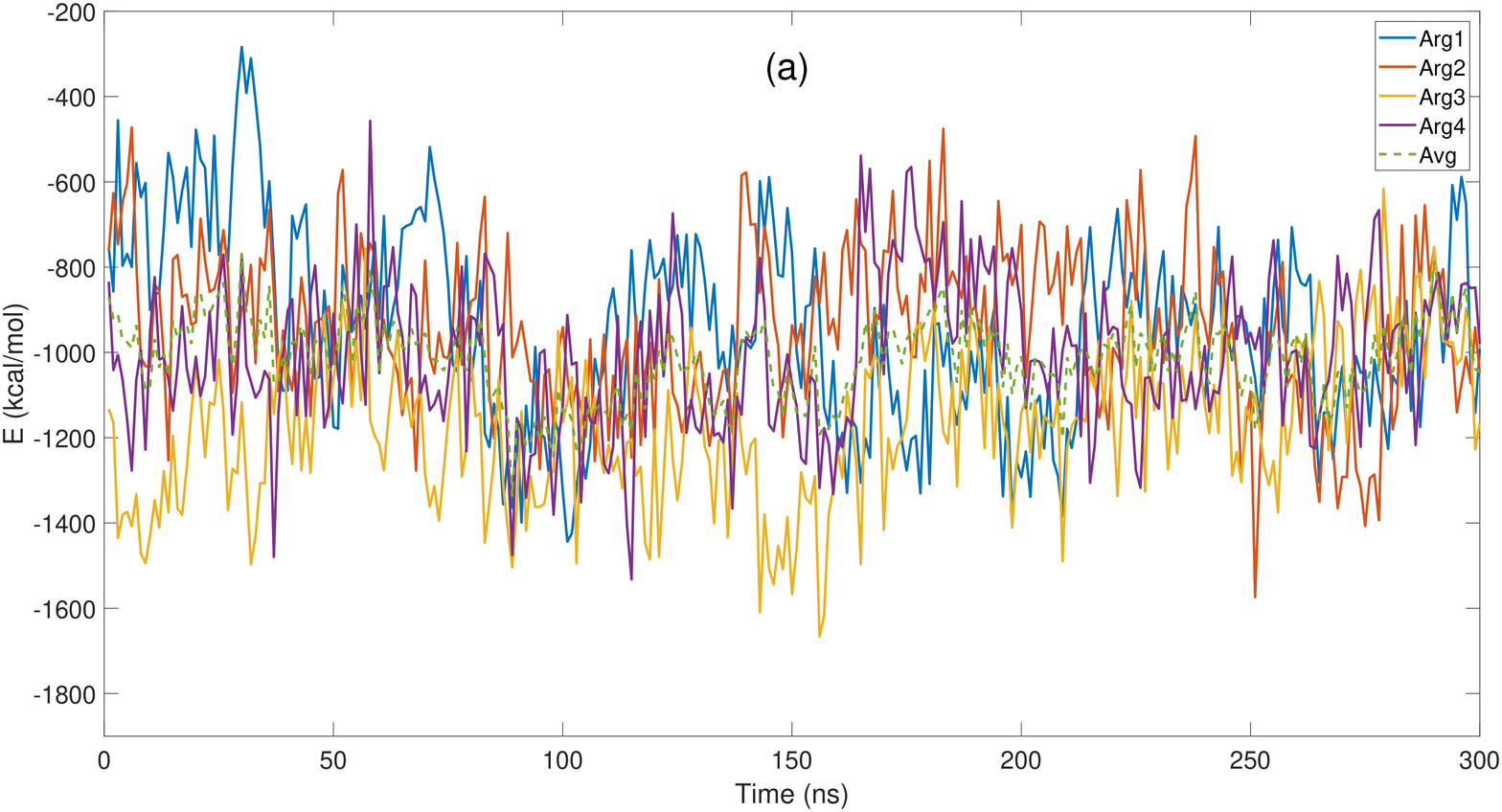}
\includegraphics[width=8cm, height=6cm]{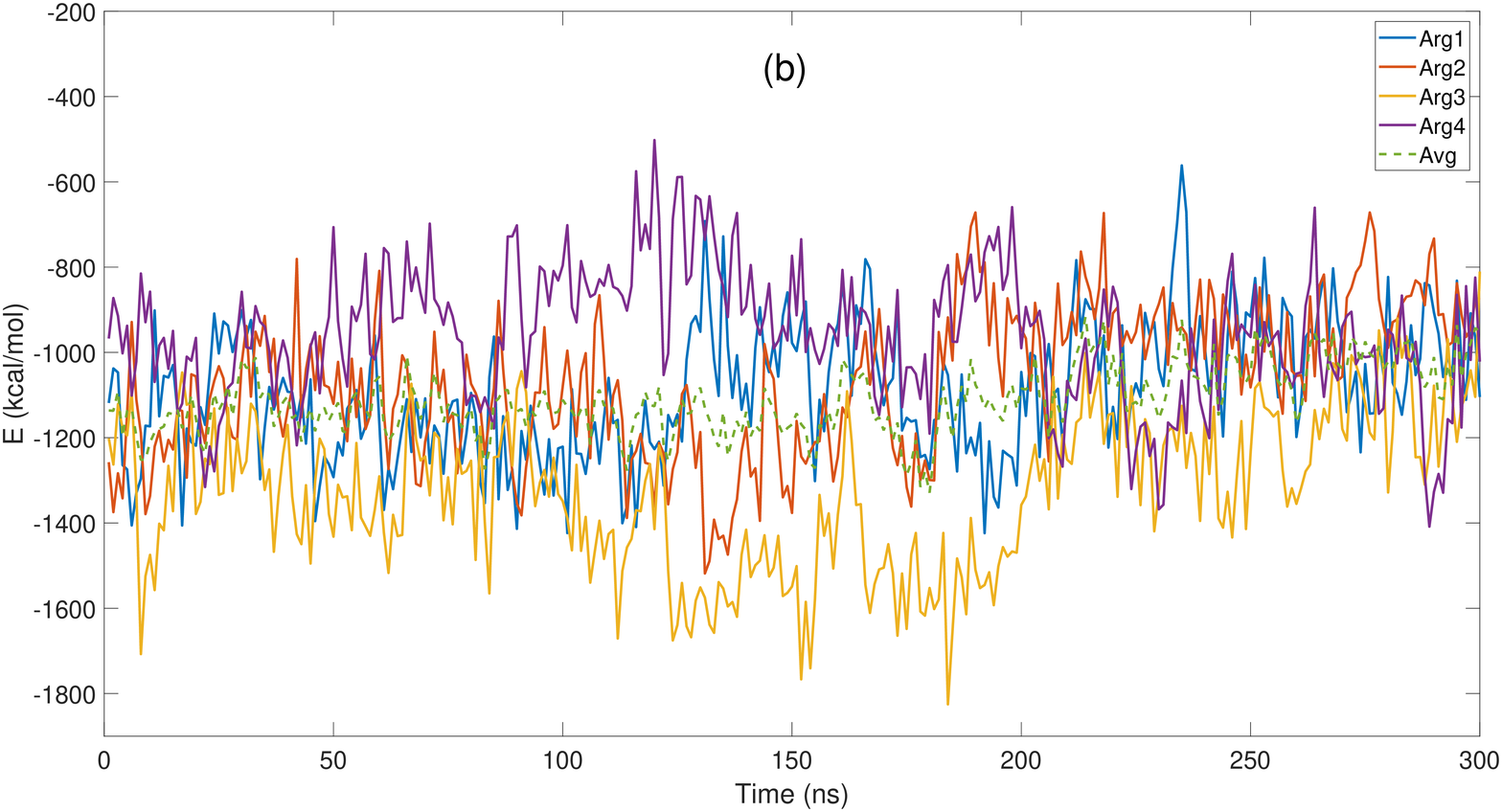}

\caption{Electrostatic energy between Arg$_9$s and the DOPC/DOPG lipids at (a) T = 310 K (b) T = 288 K.
The same color code in Fig. \ref{fig:com} is used. Avg(green dotted line) means an average value over four
Arg$_9$s. }
\label{fig:eenergy}
\end{figure}

We calculate electrostatic energy between each Arg$_9$ and the DOPC/DOPG mixture at both temperatures
using the NAMD Energy plugin in VMD. We want to see if there is any correlation between the strength of
electrostatic energy and the penetration depth. Fig. \ref{fig:eenergy}(a) shows electrostatic energy
between each Arg$_9$ and the DOPC/DOPG mixture at T = 310 K. Each color code(the same color scheme as in
the previous figures) represents the interaction between one of Arg$_9$s and the lipids mixture. Avg
denotes an average energy over all four Arg$_9$s. Fig. \ref{fig:eenergy}(b) shows the same energy, but at
T = 288 K. The average electrostatic energy at T = 288 K is slightly lower(more stable) than that at T =
310 K by 100 kcal/mol. This can be explained by comparing the average penetration depths in Figs.
\ref{fig:com}(a) and \ref{fig:com}(b). Arg$_9$ can penetrate slightly more at T = 288 K because of a
lower electrostatic energy. Note that in Fig. \ref{fig:eenergy}(b) one of the lowest energy
regions(170-200 ns) corresponds to a region which shows the largest penetration depth in Fig.
\ref{fig:com}(b). One can see a similar behavior in the case of Arg1(blue) at about 200 ns in Figs.
\ref{fig:com}(b) and \ref{fig:eenergy}(b). In both Figs. \ref{fig:eenergy}(a) and \ref{fig:eenergy}(b)
electrostatic energy between Arg3(yellow) and the lipids mixture increases(less stable) at the end of
simulation because the penetration depth of Arg3 decreases as shown in Figs. \ref{fig:com}(a) and
\ref{fig:com}(b), respectively.  Our simulations clearly suggest that electrostatic energy plays a key
role to control the penetration depth of Arg$_9$.

Then, the next question will be whether Arg$_9$ can make a pore or a prepore in the presence of strong
electrostatic energy between Arg$_9$s and the lipids mixture. At this stage it is not clear how this can
happen, but we think that Fig. \ref{fig:K288_38468_snapshots} can give us a clue. In Fig.
\ref{fig:K288_38468_snapshots} one can observe a quite number of water molecules coordinated by Arg3 when
it makes the largest penetration through the lipids mixture at T = 288 K. It was known by molecular
dynamics simulations that water molecules can enhance a membrane bending and disorder the lipid molecules.
\cite{Freites2005,Dorairaj2007,MacCallum2007} We conjecture that the concentration of Arg$_9$s will be one
of key factors to coordinate more water molecules. More water molecules will accelerate the membrane
bending and thinning, and thus there may be more interactions between Arg$_9$s and the lipids molecules in
the lower layer. However, the concentration of Arg$_9$s in our simulations seems not large enough to see
this behavior.
In the Appendix we present the membrane thinning due to the penetration of Arg3. It is obvious that the
increase of penetration of Arg$_9$s results in the membrane thinning. The detailed method to obtain the
distance between the layers is explained in the Appendix. More on the membrane bending is given in the
below.

\subsection*{Bending rigidity of the membrane is decreased due to the presence of Arg$_9$s}

It was shown that there is a correlation between CPP hydropathy and the penetration of water molecules in
the lipid bilayer and the presence of water molecules locally destabilizes the lipid order and reduces a
bending modulus of the membrane. \cite{Grasso2018} The reduction in membrane bending rigidity has also
been reported in other literature.\cite{Crosio2019,Ruseska2020}

We compare bending moduli of the DOPC/DOPG mixture at two different temperatures(T = 288 K and 310 K). We
compute the bending moduli using the method described in Ref.
\onlinecite{Khelashvili2010,Khelashvili2013,Grasso2018}.
A similar approach to calculate the bending moduli of various lipid molecules was suggested by Levine et
al. \cite{Levine2014} The authors also tried Khelashvili et al.'s method \cite{Khelashvili2013} to analyze
their data and found that their bending moduli are consistent with Khelashvili et al.'s.
We use Khelashvili et al.'s method for obtaining the bending modulus of the membrane and this will be
enough to see the temperature dependence of the bending moduli in our simulations.

It was shown that one can calculate the splay modulus $\chi$ by calculating splay angles($\alpha$) between
lipids molecules.
We use the same definition of a splay angle in Ref.
\onlinecite{Khelashvili2010,Khelashvili2013,Grasso2018}(see more in the Appendix).
There are two criteria to analyze the splay angles as explained in the previous work
\cite{Khelashvili2013,Levine2014}: 1) Only lipid pairs separated by less than 10 {\AA} are considered. 2)
At least one director should be oriented less than or equal to 10 degrees from the bilayer normal.
In this approach a normalized probability distribution is used to calculate the potential of mean
force(PMF). PMF is defined by
\begin{equation}
{\rm PMF}(\alpha) = - k_B {\rm T ~ln} ~\frac{P(\alpha)}{P_0(\alpha)}
\label{eq:Palpha}
\end{equation}
where $P_0(\alpha) = {\rm sin}(\alpha)$ is the probability distribution of a hypothetical non-interacting
particle system, $k_B$ the Boltzmann factor and T the temperature. The splay modulus $\chi$ is obtained by
a quadratic fit of the PMF data, and it corresponds to the bending modulus of a monolayer $K_m$.
The presence of Arg$_9$, e.g, in the upper layer, also affects the bending modulus of the lower layer.
Therefore, we compute the bending modulus $K_c$ of a bilayer by averaging two moduli from each layer.

\begin{figure}[ht]
\centering

\includegraphics[width=8cm, height=6cm]{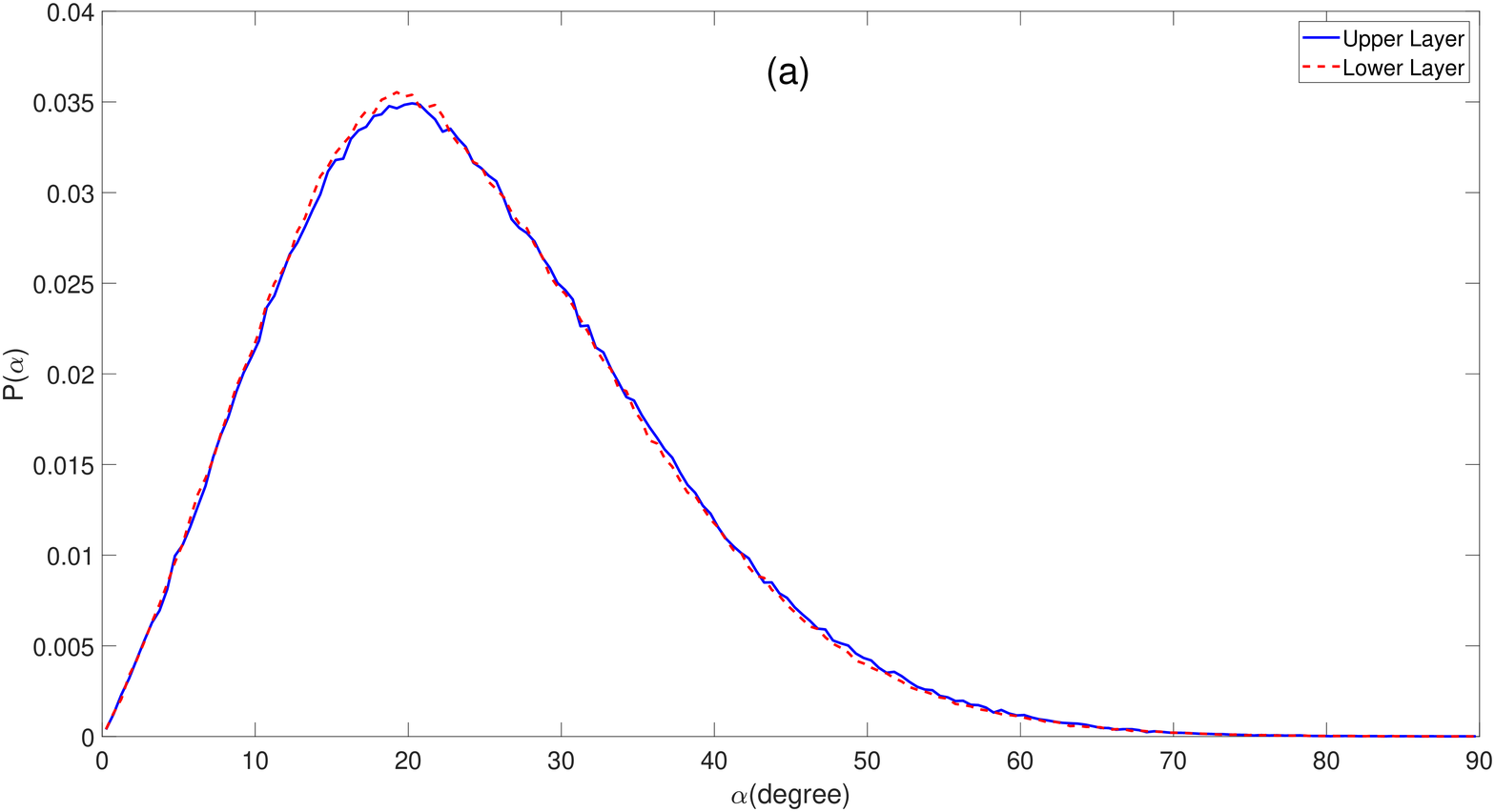}
\includegraphics[width=8cm, height=6cm]{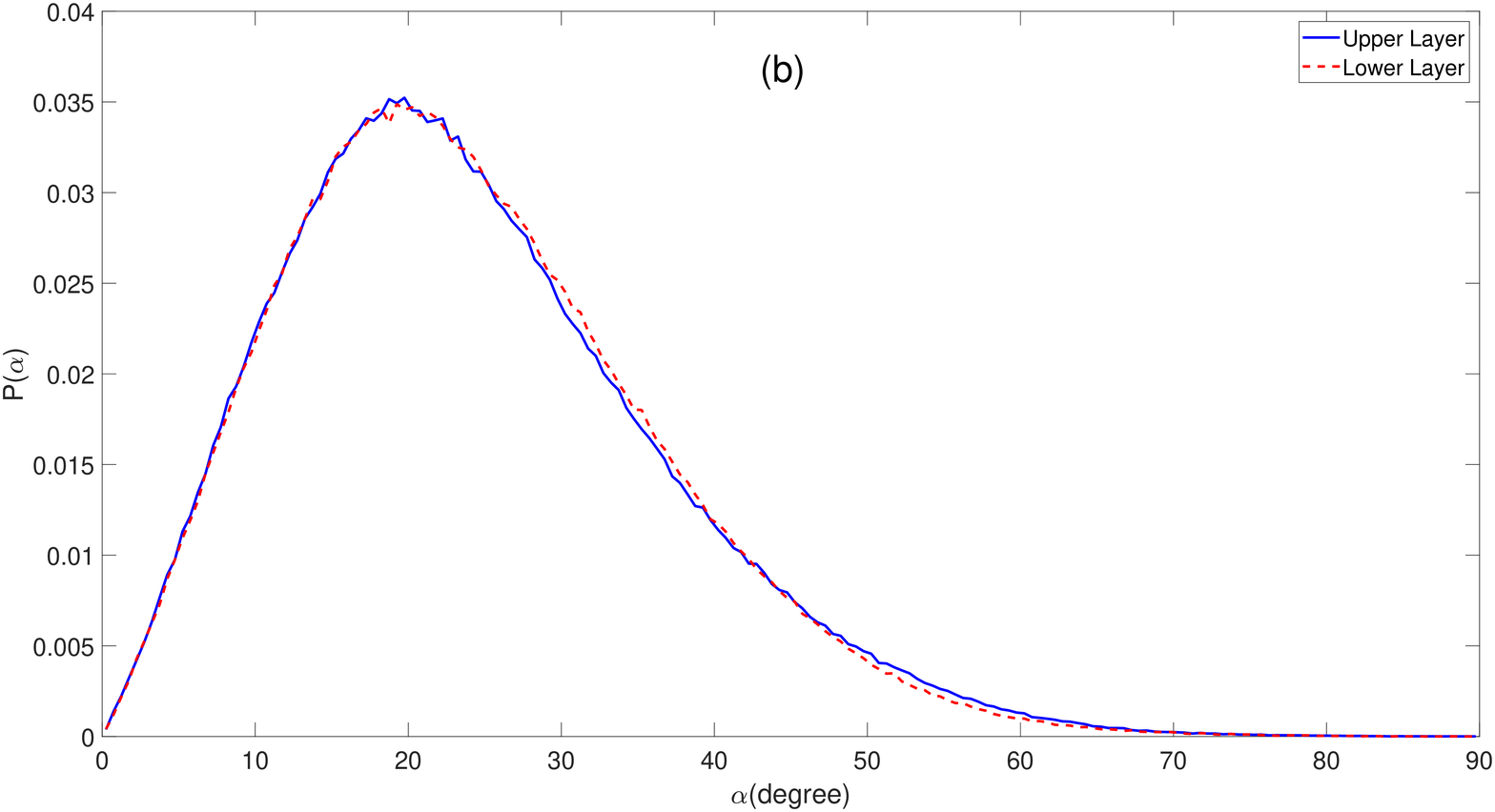}

\caption{Normalized probability distributions of $\alpha$ at (a) T = 310 K (b) T = 288 K. The blue solid
line denotes a distribution from the upper layer while the red dotted line from the lower layer.}
\label{fig:Palpha}
\end{figure}

Fig. \ref{fig:Palpha}(a) shows normalized probability distributions of $\alpha$, P($\alpha$), at T = 310
K. The blue solid line is a distribution from the upper layer, while the red dotted line comes from the
lower layer. Two lines are almost overlapped with each other and the same behavior is seen in probability
distributions at T = 288 K(see Fig. \ref{fig:Palpha}(b)). From the second order polynomial fitting of
Eq.(\ref{eq:Palpha}) as described in Ref. \onlinecite{Khelashvili2013,Levine2014,Grasso2018}, we obtain
K$_c$(310 K) = 7.6 $\times$ 10$^{-20}$ J and K$_c$(288 K) = 6.8 $\times$ 10$^{-20}$ J as bending moduli of
the DOPC/DOPG mixture(see Figs. \ref{fig:pmf_fit_310K} and \ref{fig:pmf_fit_288K} for the quadratic fits).
The bending modulus at T = 288 K is slightly smaller, but the difference is minimal. Our values( 17 $\sim$
18 $k_B$T) are slightly smaller than an experimental value, 20 $k_B$T $\pm$ 1
$k_B$T.\cite{Steinkuhler2018} This is due to the penetration of water molecules along with Arg$_9$s and
water molecules play a role to make softening of the lipids mixture.

\subsection*{A collective behavior between Arg$_9$s enhances the penetration and the membrane bending}

Li et al. \cite{Li2013} found that it is difficult for a single peptide to penetrate through the membrane
while multiple peptides can translocate across membranes by making a pore.
Hu et al. \cite{Hu2015} also showed that the water pore formation is facilitated by the aggregated charge
peptides, and the authors claimed that the cooperative effect from global and
local disturbance of membrane by random and aggregated peptides results in the pore formation.

Although we don't see any translocation of Arg$_9$s during our simulations, our simulation results are
consistent with above findings that the collective behavior between Arg$_9$s significantly changes a local
membrane curvature and results in more penetration into the lipids mixture.
In Fig. \ref{fig:K288_38468_curvature} one can see that two Arg$_9$s(red and yellow) come closer and make
a significant change in the membrane bending and thinning(see also the Appendix and Fig.
\ref{fig:membrane_thinning} for the membrane thinning).

\begin{figure}[ht]
\centering

\includegraphics[width=8cm, height=6cm]{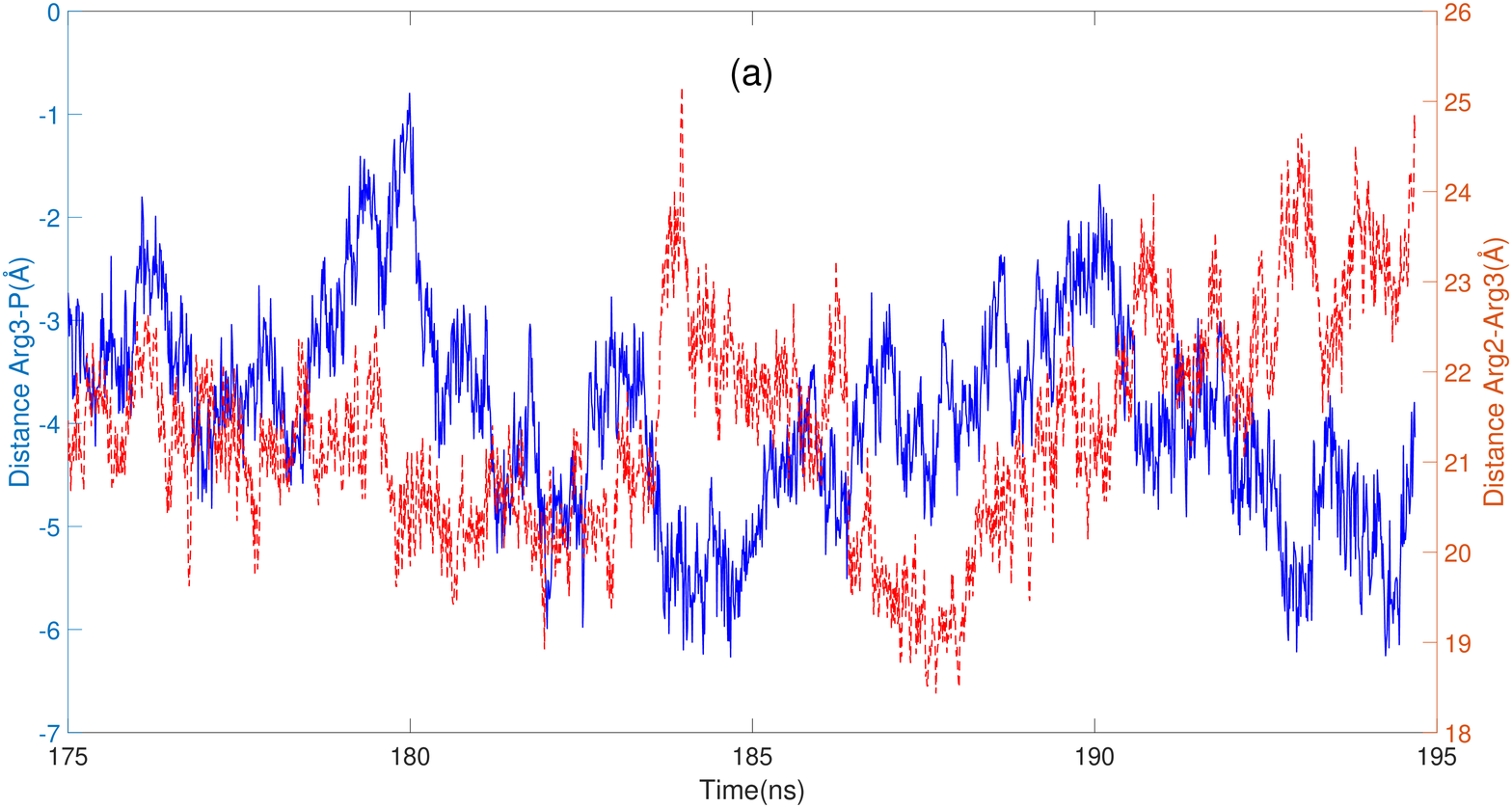}
\includegraphics[width=8cm, height=6cm]{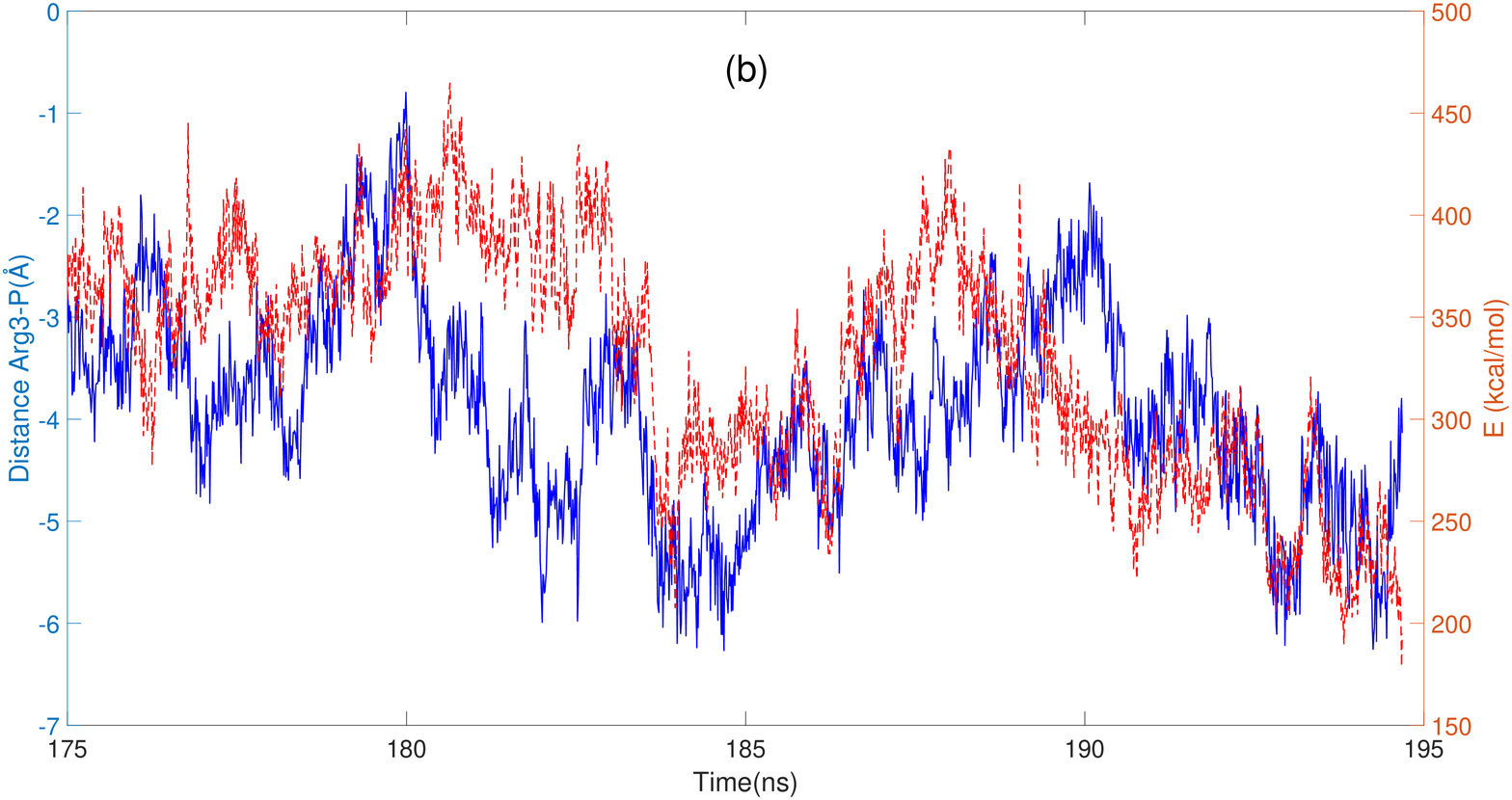}

\caption{The penetration depth of Arg3(solid blue line, left axis) at T = 288 K and (a) the distance
between Arg2 and Arg3(dotted red line, right axis) (b) electrostatic energy between Arg2 and Arg3(dotted
red line, right axis), respectively.}
\label{fig:comParg3_correlation}
\end{figure}

First, we investigate a relation between the penetration depth of Arg3(yellow) and the distance between
the center of mass of Arg2(red) and that of Arg3(yellow). In Fig. \ref{fig:comParg3_correlation}(a) the solid
blue line denotes the penetration depth of Arg3 and the dotted red line shows the distance between Arg2
and Arg3. We show only data from 175 ns to 195 ns, which include the moment of the largest penetration
depth of Arg3. Whenever Arg3 is trying to reach one of the largest penetration depths(e.g., about -6{\AA}
at 184 ns or at 193 ns), the distance between Arg2 and Arg3 is going to increase rapidly.

We have another plot which shows a comparison between the penetration depth of Arg3 and electrostatic
energy between Arg2 and Arg3. We expect that the finding in Fig. \ref{fig:comParg3_correlation}(a) regarding the
distance between Arg2 and Arg3 could be related to electrostatic energy between Arg2 and Arg3, and this
analysis will give an idea why the distance between Arg2 and Arg3 is slightly reduced when Arg3 makes the
maximum penetration depth at 184.68 ns in Fig. \ref{fig:comParg3_correlation}(a). In Fig.
\ref{fig:comParg3_correlation}(b) the solid blue line is the same as in Fig. \ref{fig:comParg3_correlation}(a)(the
penetration depth) and the dotted red line is electrostatic energy between Arg2 and Arg3. Whenever Arg3
reaches the maximum penetration, electrostatic energy between Arg2 and Arg3 becomes the minimum(or less
repulsive).

The induction of curvature by association of proteins to the
membrane has been well known for the so-called curvature sensing proteins. \cite{Maniti2014} For example,
it was known that if one protein induces a given membrane curvature, it attracts more proteins that favor
a similar curvature. \cite{Sens08} It was also shown that curvature-inducing proteins experience
attractive short-range interactions due to a membrane curvature, and this curvature-mediated interaction
leads to protein aggregation. \cite{Reynwar07}
In order to estimate membrane-mediated interactions(attractive or repulsive potential) between Arg$_9$s we
need to consider long-range interactions such as entropic membrane fluctuations, \cite{Goulian93,
Agrawal16} however, we want to focus primarily on a cooperative effect between Arg$_9$s which affects the
penetration depth and the membrane bending. Our electrostatic energy calculations seem to be enough to
show this feature.

\begin{figure}[ht]
\centering

\includegraphics[width=8cm, height=6cm]{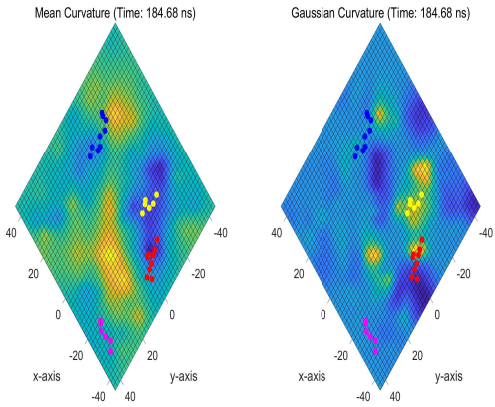}

\caption{The mean curvature(left) and the Gaussian curvature(right) of the membrane in the presence of
Arg$_9$s(showing only C$_\alpha$s) when Arg3(yellow) makes the largest penetration at T = 288 K. The blue
color in each curvature means a negative value while the yellow a positive.}
\label{fig:curvature}
\end{figure}

We want to briefly mention the usefulness of membrane curvatures such as the mean curvature($H$) or the
Gaussian curvature($K$) in our simulations. \cite{Schmidt2010} Each curvature is defined by
\begin{eqnarray}
 H &=& \frac{1}{2} (c_1 + c_2)
 \\ \nonumber
K &=& c_1  c_2
\label{eq:curvature}
\end{eqnarray}
where $c_1$ and $c_2$ are the principal curvatures at a point on the membrane.
Fig. \ref{fig:curvature} represents the mean curvature and the Gaussian curvature of the membrane in the
presence of Arg$_9$s when Arg$_3$ shows the largest penetration at T = 288 K(see also Fig.
\ref{fig:K288_38468_curvature}). The blue color in the figure means a negative curvature while the yellow
denotes a positive curvature. As we expect the mean curvature near Arg2(red) and Arg3(yellow) is negative
which means a concave deformation. We have checked that the Gaussian curvature near Arg3 is always
positive during the simulation at T = 288 K, which means no evidence of the pore formation.
\cite{Schmidt2010}

We expect that both the concentration of Arg$_9$s and the collective behavior are important in
translocation. More simulation studies with different concentration of Arg$_9$s may be necessary to gain
more insights into the collective behavior.

\section{Discussion}

Our simulations show that Arg$_9$ can penetrate into the DOPC/DOPG mixture, but it can not pass through
the mixture both at T = 310 K and at T = 288 K. It doesn't show any pore or prepore as suggested in Refs.
\onlinecite{Herce2009, Islam2018}. Although we don't see any translocation of Arg$_9$, our simulations
suggest that electrostatic energy between Arg$_9$ and the lipids mixture plays a role in determining the
penetration depth. A few remarks regarding our results are given below.

The role of the Arg side chain was identified in the previous MD simulations.\cite{Herce2009}
The Arg side chains bind to the phosphate groups of the phospholipids, then this binding makes large distortions of
the bilayer which lead to the formation of a pore.

Our simulation shows that more than 70 water molecules also penetrate into the lipids mixture when Arg$_9$
reaches the maximum penetration depth(see Fig. \ref{fig:K288_num_water}), however, Arg$_9$s in our
simulations don't make such strong distortions in the DOPC/DOPG mixture which can lead to a pore.
We expect that the translocation may be difficult at low concentration of Arg$_9$s due to the attractive
potential energy between Arg$_9$s(positive charge) and DOPG lipids(negative charge). We find that all
Arg$_9$s were strongly bound to the DOPG lipids during the simulations. We conjecture that the ability of
the translocation can be highly dependent on the concentration of Arg$_9$s. To our knowledge, it has not
yet known how positively charged Arg$_9$s pass through the DOPC/DOPG mixture within all-atom MD
simulations.
One of crucial reasons will be a simulation timescale. It was claimed that spontaneous translocation is
not expected within timescales accessible to MD simulations, unless nonphysical restraints
are applied to accelerate the underlying process.\cite{Yesylevskyy2009}
In that case, a path sampling approach(e.g., the weighted ensemble method \cite{westpa}) might be one of
solutions to reveal the underlying mechanism of the translocation of Arg$_9$s.

It was known that charged surfaces are more rigid than neutral ones, therefore, the decrease
in membrane rigidity could be explained by considering that the arginine-rich peptide partially
neutralizes the membranes.\cite{Crosio2019}  Crosio et al. \cite{Crosio2019} claimed based on their
experiments in the DOPC/DOPG(1:1) environment that the recruiting of the anionic DOPG lipids in the regions
close to the arginine-rich peptide would lead to an enrichment of DOPC lipids in other regions, subsequently decreasing the
observed bending rigidity.
The incorporation of water coupled with peptide translocation and local
membrane thinning can be another explanation for the reduced bending rigidity of the membrane.
\cite{Grasso2018} Our simulations suggest that both the partial neutralization and the membrane thinning
affect a softening of the DOPC/DOPG mixture. As we mentioned above, Arg$_9$s are strongly bound to DOPG
lipids during the simulations and this results in the partial neutralization of the membrane. We also
showed the membrane thinning due to the penetration of water molecules in the previous section. It will be
interesting to investigate which factor(the neutralization or the membrane thinning) can contribute more
on the softening of the membrane. This can be done using different concentration of Arg$_9$s as well as
using different DOPC/DOPG ratios.

\begin{figure}[ht]
\centering

\includegraphics[width=8cm, height=6cm]{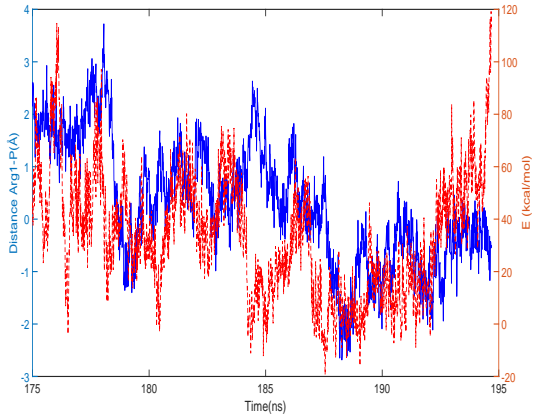}

\caption{The penetration depth of Arg1(solid blue line, left axis) at T = 288 K and the sum of
electrostatic energy between Arg1 and Arg2, and that between Arg1 and Arg3(dotted red line, right axis)}
\label{fig:comParg1_eenergyPPPQPR}

\end{figure}

Although the total number of Arg$_9$s and the system size may not be large enough to clearly show the
collective behavior in our simulations, our results suggest that it can affect the membrane bending and
thinning and thus the penetration of Arg$_9$s.
Movie S1 shows how each Arg$_9$ makes membrane deformations during the time frame between 175 ns and 195
ns which includes the moment that Arg3 shows the largest penetration depth at about 185 ns.
One can see in this movie that there are two main local minima made by Arg1(blue) and Arg3(yellow).
Arg2(red) is already got closer to Arg3(yellow) and it is trapped in a local minimum during this time
frame. One interesting finding is that sometimes two local minima are connected to each other, and this
results in a drastic change in the membrane bending(see, e.g., at 188.6 ns, 190.4 ns, and 191.6 ns in the
movie). One can find an increased penetration depth and a drastic decrease of electrostatic energy of
Arg1(blue) at about 190 $\sim$ 200 ns in Figs. \ref{fig:com}(b) and \ref{fig:eenergy}(b), respectively.
We compute electrostatic energy between Arg1(blue), Arg2(red), and Arg3(yellow) as we did previously for
the interaction between Arg2 and Arg3. Fig. \ref{fig:comParg1_eenergyPPPQPR} represents these calculations
with the penetration depth of Arg1 for comparison. The blue solid line is the penetration depth of Arg1
and the red dotted line is the sum of two electrostatic energy: One is electrostatic energy between Arg1
and Arg2, and the other one is that between Arg1 and Arg3. There is an interesting energy drop at about
185 ns, and this time frame corresponds to the moment that Arg3 makes the maximum penetration depth. One
can see that the penetration increases at the time frames mentioned above, i.e., at 188.6 ns, 190.4 ns,
and 191.6 ns, and electrostatic energy becomes minimum (or less repulsive) at these time frames.
It shows that there exits the collective behavior between Arg$_9$s which enhances the disorder of the
membrane.
Further simulations with more detailed curvature analyses \cite{Schmidt2010,Maniti2014} will be helpful to
identify the factors(e.g., concentration of Arg$_9$s, a distribution of DOPG lipids in the DOPC/DOPG
mixture, etc.) which control the collective behavior between Arg$_9$s and their effects on the membrane
bending.

\section{Conclusion}

The uptake mechanisms of CPPs have been extensively studied in both experiments and computer simulations.
However, they are not still conclusive. The difficulty comes from the fact that the uptake pathway is not
the same for different families of CPPs and depends on the experimental conditions.\cite{Madani2011} There
is no doubt that an MD simulation is still one of valuable tools to reveal functional and mechanical
properties of different families of CPPs and their interactions with various types of lipid membranes.

Our simulations suggest that the electrostatic interaction between Arg$_9$s and the lipids molecules is
important at the early stage of the penetration, and water molecules coordinated by Arg$_9$s make a
membrane thinning which results in a softening of the bending rigidity of the membrane. The collective
behavior between Arg$_9$s also plays a role to help the penetration of Arg$_9$s into the membrane.
Although we don't see any direct translocation of Arg$_9$ across the DOPC/DOPG bilayer in our simulations,
our analyses will be helpful for understanding functional properties of other CPPs and their interactions
with various membranes using MD simulations.

\section*{Supplementary Material}

See supplementary material for the supplemental movie S1.

\begin{acknowledgments}
The author thanks Hanna Salman and Xiao-lun Wu for valuable comments on the early stage of this work, and
he also thanks the University of Pittsburgh for support during his stay. This work was supported in part
by the start-up fund from the Daegu Gyeongbuk Institute of Science \& Technology(DGIST).
\end{acknowledgments}

\section*{DATA AVAILABLE STATEMENT}
The data that support the findings of this study are available from the corresponding author upon
reasonable request.


\appendix

\section{\label{sec_app1}Membrane thinning due to the penetration of Arg$_9$}

\begin{figure}[ht]
\centering

\includegraphics[width=8cm, height=7cm]{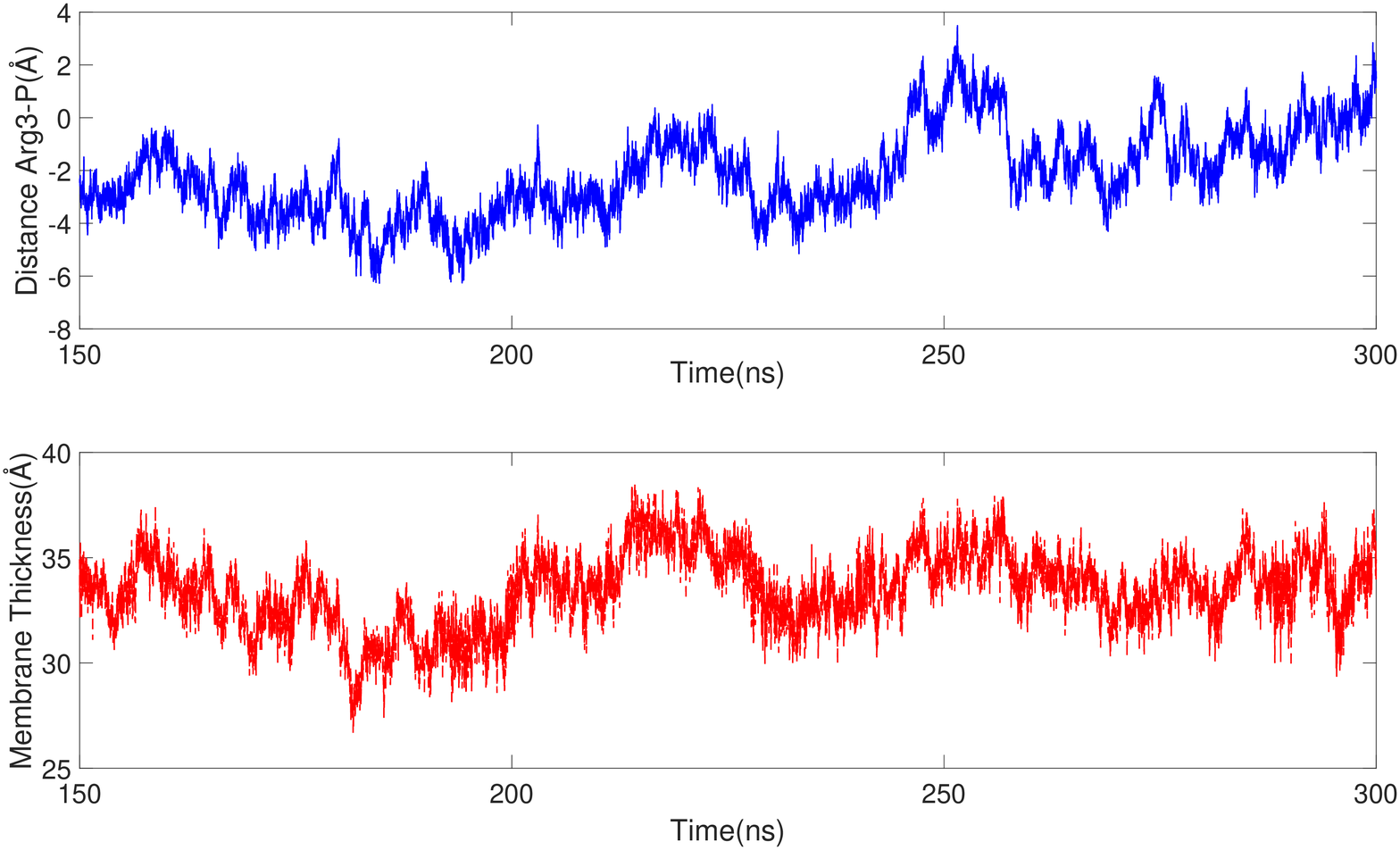}

\caption{The penetration depth of Arg3(the upper panel)  and the membrane thickness(the lower panel) at T
= 288 K. }
\label{fig:membrane_thinning}
\end{figure}

We investigate a relation between the penetration depth of Arg3(one of Arg9 peptides in our simulations)
and the membrane thickness at T = 288 K(see Fig. \ref{fig:membrane_thinning}). When the penetration depth
of Arg3 increases, the membrane thickness near Arg3 is expected to be decreased. The first panel
represents the penetration depth of Arg3(the distance between the center of mass of Arg3 and the center of
masses of phosphorus(P) atoms in the upper leaflet, where a leaflet is a surface which consists of 95
phosphorus(P) atoms in each layer) between 150 ns and 300 ns. The negative value means that the center of
mass of Arg3 is below than the center of masses of phosphorus(P) atoms, and the positive means the
reverse. The second panel shows time variations of the membrane thickness due to the penetration of Arg3.
The smallest thickness is about 27 {\AA}, which is almost 10 {\AA} smaller than that of a membrane without
any interactions with Arg9s(a rest state). We compute the membrane thickness as follows: First, we find
the three nearest phosphorus(P) atoms in the upper leaflet from the center of mass of Arg3 and calculate
an average z position of these three atoms. Next, we do the same thing in the lower leaflet, namely, we
obtain another average z position of three nearest phosphorus(P) atoms in the lower leaflet. Then, we
define a membrane thickness as a distance between those two z positions. One can see that two figures show
a very similar trend. This is what we exactly expected. We choose Arg3 because it shows the largest
penetration depth at T = 288 K. The other Arg9s can have different patterns in the membrane thickness.

\section{\label{sec_app2}Bending modulus of the DOPC/DOPG mixture}

\begin{figure*}[ht]
\centering

\includegraphics[width=15cm, height=6cm]{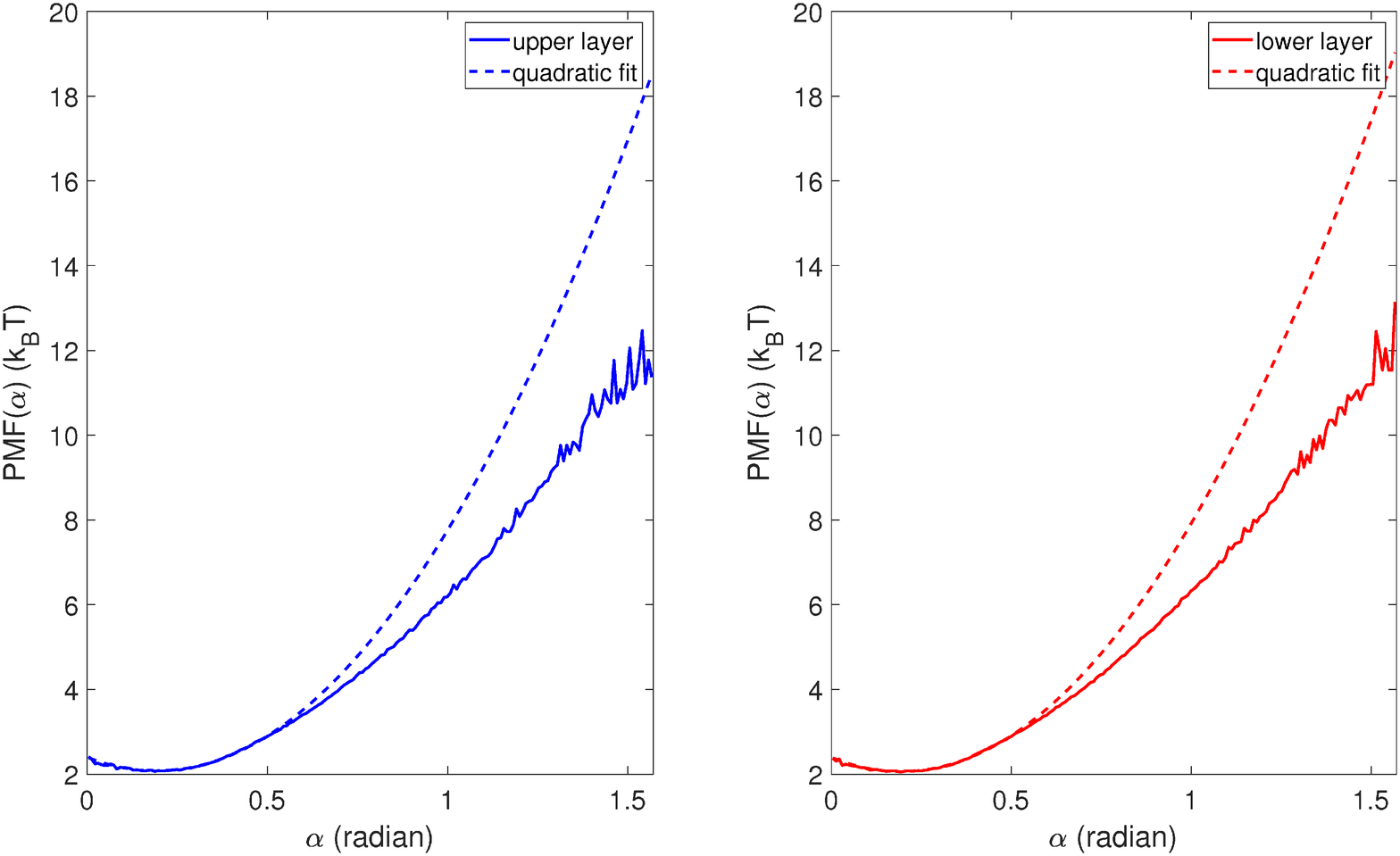}

\caption{PMF profiles and quadratic fits for both the upper and the lower layers at T = 310 K. The blue
solid line and the blue dotted lines in the left figure are a profile from the upper layer and a quadratic
fit at [10, 30] degrees, respectively. The right figure is the same, but from the lower layer.}
\label{fig:pmf_fit_310K}
\end{figure*}

One of mechanical properties of a membrane is its bending modulus. It was known that the penetration of a
quite amount of water molecules plays a role to decrease a bending rigidity of a membrane. We investigate
the bending modulus of the DOPC/DOPG mixture in the presence of Arg9s both at T = 310 K and at T = 288 K.
We use Khelashvili et al.’s approach \cite{Khelashvili2013} to get the bending modulus from MD
simulations. One can compute a bending modulus from a probability distribution of splay angles. Here is a
general method to obtain splay angles.\cite{Khelashvili2013} First, local lipid director vectors $\vec{v}$
is defined, where $\vec{v}$ is the vector that connects the midpoint between the phosphate and backbone C2
atoms to the center of mass of the three terminal carbons on the two lipid chains.
The splay
modulus for a particular pair of lipid molecules is calculated by defining a splay angle ($\alpha$)
between $\vec{v}$  vectors and constructing probability distributions P($\alpha$) (see the main text and
Ref. \onlinecite{Khelashvili2013}). Fig. \ref{fig:pmf_fit_310K} represents the potential of mean force
(PMF) profile PMF($\alpha$)= - $k_B$ T ln (P($\alpha$)/sin($\alpha$)) and a quadratic fit at [10, 30]
degrees at T = 310 K, where sin ($\alpha$) is the probability distribution of a hypothetical
non-interacting particle system. A figure in the left represents the PMF profile(the blue solid line) from
the upper layer and a quadratic fit(the blue dotted line). A figure in the right shows the same plots from
the lower layer.

\begin{figure*}[ht]
\centering

\includegraphics[width=15cm, height=6cm]{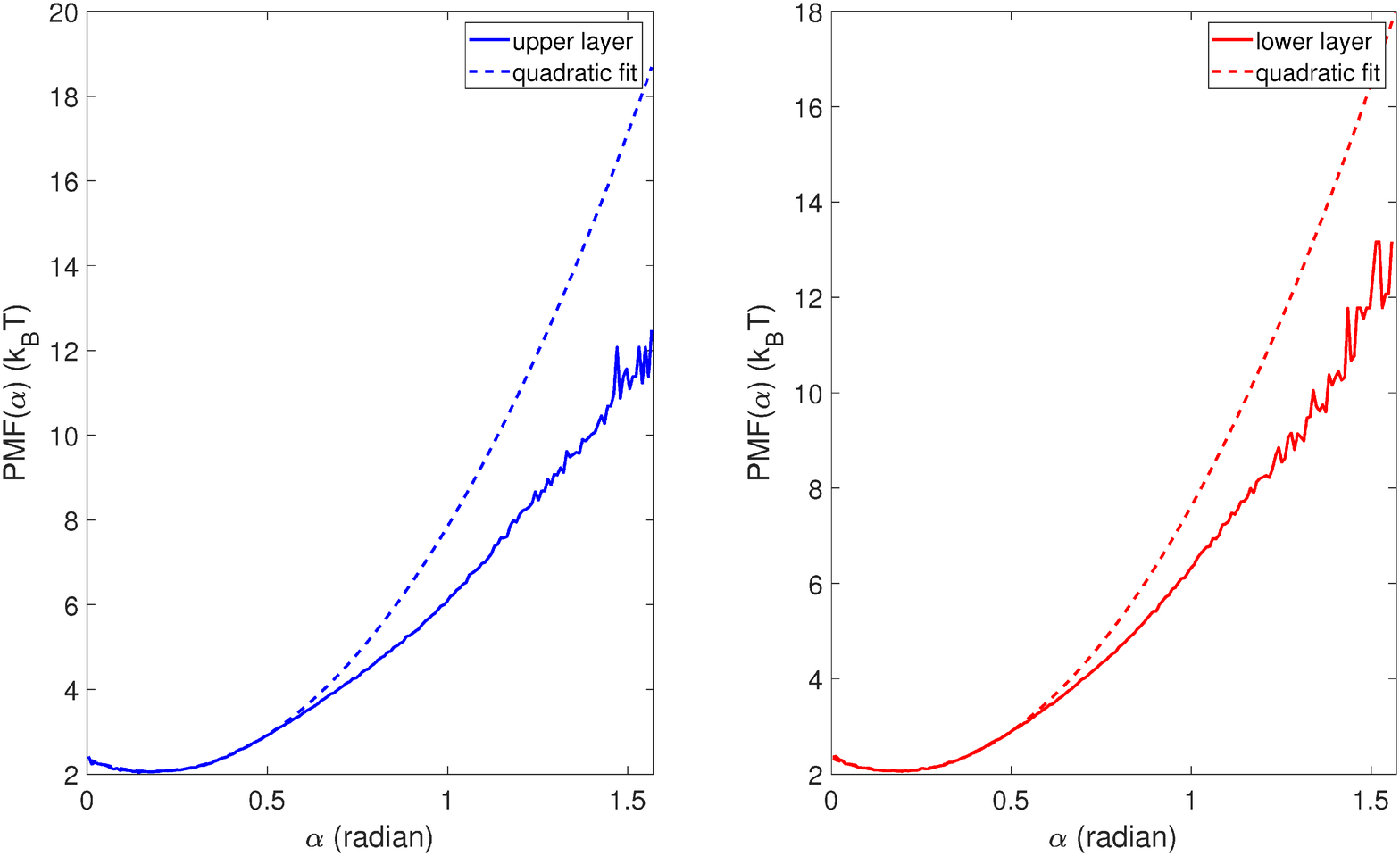}

\caption{The same figure as in Fig. \ref{fig:pmf_fit_310K}, but at T = 288 K.}
\label{fig:pmf_fit_288K}
\end{figure*}

In order to obtain a quadratic fit of the PMF profile we use the Curve Fitting Tool of MATLAB. We obtain
the following bending modulus at T = 310 K: 17.408 $k_B$ T(the upper layer) and 17.990 $k_B$ T(the lower
layer). We take an average of these two values, 17.699 $k_B$ T , as the bending modulus of the DOPC/DOPG
mixture at T = 310 K. Fig. \ref{fig:pmf_fit_288K} is the same as in Fig. \ref{fig:pmf_fit_310K}, but at T
= 288 K. As for the probability distributions of the splay angles, there are no significant changes at T =
288 K. Using the same approach, we get the following bending modulus: 17.428 $k_B$ T(the upper layer) and
16.600 $k_B$ T(the lower layer). Again, we take an average of these two values, 17.014 $k_B$ T, as the
bending modulus of the DOPC/DOPG mixture at T = 288 K.


\bibliography{cpp}

\end{document}